\documentclass[twocolumn,secnumarabic,amssymb, nobibnotes, aps, prd]{revtex4}
\usepackage{graphicx}

\begin{document}
\title{Quantum periodicity in the critical current of superconducting rings with asymmetric link-up of current leads}
\author{A. A. Burlakov, A. V. Chernykh, V.L. Gurtovoi, A. I. Ilin, G. M. Mikhailov, A.V. Nikulov and V.A. Tulin}
\affiliation{Institute of Microelectronics Technology and High Purity Materials, Russian Academy of Sciences, 142432 Chernogolovka, Moscow District, RUSSIA.} 
\begin{abstract} We use superconducting rings with asymmetric link-up of current leads for experimental investigation of winding number change at magnetic field corresponding to the half of the flux quantum inside the ring. According to the conventional theory, the critical current of such rings should change by jump due to this change. Experimental data obtained at measurements of aluminum rings agree with theoretical prediction in magnetic flux region close to integer numbers of the flux quantum and disagree in the region close to the half of the one, where a smooth change is observed instead of the jump. First measurements of tantalum ring give a hope for the jump. Investigation of this problem may have both fundamental and practical importance. 
\end{abstract}

\maketitle 

\narrowtext

\section{Introduction}
\label{}
The observation of quantum periodicity in the transition temperature of a superconducting cylinder by W.A. Little and R.D. Parks \cite{LP1962} has demonstrated macroscopic quantum phenomenon. According to the conventional explanation \cite{Tinkham} the Little-Parks effect \cite{LP1962} is observed due to the discreteness of the permitted states of superconducting pairs in a cylinder or ring. According to the Bohr-Sommerfeld quantization 
$$\oint_{l}dl p  = \oint_{l}dl (mv + qA) = n2\pi \hbar = nq\Phi_{0} \eqno{(1)}$$ 
the energy difference $E_{n+1} - E_{n} = mv_{n+1}^{2}/2 - mv_{n}^{2}/2 = (2n+1)\hbar ^{2}/2mr^{2}$ between adjacent permitted states $n + 1$ and $n$ in a ring with a macroscopic radius $r \approx 1 \ \mu m = 10^{-6} \ m$ should be much smaller than in atom with the Bohr radius $r_{B} \approx 0.05 \ nm = 5 \ 10^{-11} \ m$. Therefore the quantum phenomena, such as the persistent current of electrons, can be observed in nano-rings with a real radius $r > 300 \ nm$ only at very low temperature \cite{PCScien09,PCPRL09}. Superconductivity is the macroscopic quantum phenomena thanks to the impossibility for Cooper pairs to change their quantum state $n$ individually \cite{FPP2009}. The energy difference $E_{n+1} - E_{n}$ in superconductor ring in $N_{s} $ times higher \cite{FPP2009} then in the non-superconductor one because of the same $n\hbar $ angular momentum of all $N_{s} = Vn_{s} = 2\pi rsn_{s}$ pairs containing in the ring with the volume $V$, the radius $r$ and the section area $s$. Therefore the spectrum of the permitted states of a real superconducting ring with a macroscopic radius $r \approx 1 \ \mu m $ is strongly discrete \cite{APL2016}, almost like of atom. Some authors consider superconducting loops as artificial atoms \cite{ArtAt17,ArtAt16}.
  
Superconducting rings or loops give us an opportunity for more detailed research of phenomena connected with the Bohr quantization. We can observe quantum phenomena connected with the change of the winding number $n$, describing of the angular momentum $n\hbar $, with magnetic field $B$. The magnetic flux $\oint_{l}dl A= \Phi = BS \approx B\pi r^{2}$ increases up to the flux quantum $\Phi _{0} = 2\pi \hbar /q $, see (1), equals $\approx 4140 \ T \ nm^{2}$ for electron $q = e$ and $\approx 2070 \ T \ nm^{2}$ for Cooper pair $q = 2e$, at an inaccessibly high magnetic fields $\Phi_{0}/ \pi r_{B}^{2} \approx 530000 \ T$ inside atom orbit with $r_{B} \approx 0.05 \ nm $ and at accessibly low value  $B_{0}  = \Phi_{0}/ \pi r^{2} \approx 0.0026 \ T $ inside a superconducting ring with $r \approx 500 \ nm$. Therefore quantum periodicity in the transition temperature \cite{LP1962} and other variables can be easy observed. Moreover, a ring, unlike the atom, can be produced in various forms and we can connect the current leads to it. Measurements of the critical current \cite{JETP07J,PCJETP07} and other parameters \cite{PCJETP07,Letter2003,Letter07,Kulik75} of asymmetric superconducting rings with the help of the current leads have allowed to discover a new quantum phenomenon and to reveal some contradictions between theoretical predictions and experimental results. The rectification effect discovered due to these measurements allows to use a system with large number of asymmetric rings for ultrasensitive detection of non-equilibrium noises \cite{APL2016} and for the experimental investigation of the possibility of observing persistent voltage \cite{PLA2012}. The rings investigated formerly \cite{JETP07J,PCJETP07,Letter2003,Letter07,Kulik75,PLA2012} were asymmetric due to different cross-section of their halves. Here we present experimental results obtained at measurements of the rings with asymmetric link-up of current leads. 

\section{Periodical dependencies in magnetic field}
 \label{}
The quantum periodicity observed first by W.A. Little and R.D. Parks \cite{LP1962} is possible due to the Aharonov-Bohm effect \cite{AB1959}, i.e. the influence of the magnetic vector potential $A$ of the canonical momentum $p = mv + qA$ of a particle with charge $q$. Aharonov and Bohm have noted \cite{AB1959} that according to the canonical definition of the operator velocity $m\hat{v} = \hat{p}-qA  = -i\hbar \nabla -qA$ \cite{LLqm} the velocity $\Psi^{*} m\hat{v}\Psi =  |\Psi |^{2}mv = |\Psi |^{2}(\hbar \nabla \varphi  -qA)$  of a particle with a charge $q$ along a closed contour $l$ is connected 
$$m\oint_{l}dl v = \hbar \oint_{l}dl\nabla \varphi - q\Phi \eqno {(2)}$$ 
with the change $\oint_{l}dl\nabla \varphi $ of the wave function $\Psi = |\Psi |\exp i\varphi $ phase $\varphi $ along $l$ and the magnetic flux $\Phi = \oint_{l}dl A$ inside $l$. According to the relation (2) the magnetic flux $\Phi$ should influence on the phase change $\oint_{l}dl\nabla \varphi $ when it cannot influence on the velocity $m\oint_{l}dl v $ \cite{AB1961}. In the opposite case the velocity changes with magnetic flux according to (1) when the $\oint_{l}dl\nabla \varphi $ value cannot change. In the first case Aharonov and Bohm have predicted the quantum periodicity in the arrival probability of a charge particle at a point of detecting screen on $\oint_{l}dl A= \Phi$ in the two-slit interference experiment \cite{AB1959}. The quantum periodicity in the velocity of Cooper pairs \cite{Tinkham} is observed in superconducting ring due to the requirement that the complex wave function must be single-valued $\Psi  = |\Psi |\exp i\varphi = |\Psi |\exp i(\varphi + n2\pi)$ at any point of $l$. According to this requirement, conforming with the Bohr-Sommerfeld quantization (1), $\oint_{l}dl\nabla \varphi = n2\pi $ and the velocity of Cooper pairs should depend on the magnetic flux and the winding number $n$. 

The current $I_{p} = sq|\Psi |^{2}v = sqn _{s}v $ of $N_{s} = \oint_{l}dl sn_{s}$ Cooper pairs, called the persistent current 
$$I_{p} = qsn _{s}v = \frac{q2\pi \hbar }{lm\overline{(sn _{s})^{-1}}}(n-\frac{\Phi }{\Phi_{0}}) = \frac{n\Phi_{0} - \Phi}{L_{k}}  \eqno{(3)}$$
is observed in superconducting ring with a small cross-sectional area $s \ll \lambda _{L}^{2}$ \cite{PRB2001}, when the permitted states of the circular velocity of Cooper pairs
$$\oint_{l}dlv  = \frac{2\pi \hbar }{m}(n-\frac{\Phi }{\Phi_{0}})   \eqno{(3a)}$$
is discrete because of the quantization (1) and the state with $v = 0$ is forbidden because of the Aharonov-Bohm effect (2).  Here $\overline{(sn _{s})^{-1}} = l^{-1}\oint_{l}dl (sn _{s})^{-1}$ \cite{PRB2001} when the cross-sectional area $s$ and the density of Cooper pairs $n _{s}$ are functions of position along the circumference of the ring $l$; the London penetration depth $\lambda _{L} = (m/\mu _{0}q^{2}n_{s})^{0.5} $ is also a function of position; $L _{k} = ml\overline{(sn _{s})^{-1}}/ q^{2} = (\overline{\lambda _{L}^{2}/s}) \mu _{0}l$ is the kinetic inductance of the ring with the length $l = 2\pi r$ equal $L _{k} = ml/ q^{2}n_{s}s = (\lambda _{L}^{2}/s) \mu _{0}l$ when the cross-sectional area $s = wd$ and the density $n_{s}$ of Cooper pairs with the charge $q = 2e$ are uniform along the circumference; the total magnetic flux $\Phi = BS + L _{f}I_{p}$ is produced  by the external magnetic field $B$ inside the ring with the area $S =\pi r^{2}$ and the persistent current $I_{p}$. The energy of the magnetic field $L _{f}I_{p}^{2}/2$ of thin wire is much lower than the kinetic energy $L _{k}I_{p}^{2}/2$ \cite{Tinkham} because of the inequality $L_{f}  \approx \mu _{0}l \ll L _{k} = (\lambda _{L}^{2}/s) \mu _{0}l $ valid at $s \ll \lambda _{L}^{2}$. Therefore "{\it we can always neglect it for a sufficiently thin conductor}" \cite{Tinkham}. According to the conventional theory \cite{Tinkham} the total energy of the persistent current (3) in a ring with weak screening $s \ll \lambda _{L}^{2}$ equals approximately its kinetic energy 
$$E_{k} =  \frac{L _{k}I_{p}^{2}}{2} = \frac{I_{p}(n\Phi_{0} - \Phi )}{2} =  \frac{(n\Phi_{0} - \Phi )^{2}}{2L_{k}} \eqno{(4)}$$ 
and $\Phi = BS + L _{f}I_{p}  \approx BS$. The relations (3) and (4) are valid when the density of Cooper pairs is not zero $n _{s} > 0$ in all ring segments. The winding number $n$ becomes undefined when the density vanishes $n _{s} = 0$ at a point of the ring \cite{PLA2012QF}. In this case the persistent current (3) vanishes $I_{p} = 0$ and the spectrum of the permitted states (4) becomes continuous \cite{PLA2012QF} since $1/L _{k} = q^{2}/ml\overline{(sn _{s})^{-1}}  = 0$ at $1/\overline{(n _{s})^{-1}} = 0$. 


\section{Theoretical predictions and experimental results}
 \label{}
Experiments corroborate the change of the persistent current (3) with the external magnetic flux $B$. The persistent current in isolated flux-biased ring increases in time $dI_{p}/dt = -L _{k}^{-1}(d\Phi /dt)  \approx -L _{k}^{-1}S(dB /dt) $ with $B$ in accordance with the Newton second law $mdv/dt = qE = -(q/l) d\Phi /dt$ up to a critical value \cite{nJumpGeim,nJumpMoler,nJumpNat}. The pair velocity, equal $|v| = (\hbar/mr)|n - \Phi /\Phi _{0}|$ in a homogeneous ring with the radius $r$ according to (3a), cannot exceed the depairing velocity $v _{c} = \hbar /m \surd 3 \xi (T)$ at which the density of Cooper pairs decreases down to $n_{s} = 2n_{s0}/3$, where $n_{s0}$ is the density at $|v| = 0$ and $\xi (T) $ is the correlation length of the superconductor \cite{Tinkham}. The persistent current (3) decreases by jump \cite{nJumpGeim,nJumpMoler} at $|n - \Phi /\Phi _{0}| \approx r/\surd 3 \xi (T)$ due to the velocity (3a) jump with the change of the winding number $n$. According to the requirement $\oint_{l}dl\nabla \varphi = n2\pi $ the $n$ change occurs due to phase slip \cite{nJumpNat} when the ring or a ring segment is switched in normal state, from $n_{s} = 2n_{s0}/3$ to $n_{s} = 0$, for a while. The winding number $n$ corresponds to the minimum of the kinetic energy (4) with predominant probability $P _{n} \propto \exp{- E_{k}(n)/ k _{B}T} $ when the ring is switched in normal state by an external current \cite{JETP07J,PCJETP07,Letter2003}, noise \cite{APL2016,Letter07,PLA2012,Dubonos02}, or thermal fluctuations \cite{LP1962,Kulik75,PCScien07}. Therefore the quantum periodicity in the critical current \cite{JETP07J,PCJETP07}, in the dc voltage \cite{APL2016,Letter07,PLA2012,Dubonos02,Letter2003}, in the resistance \cite{LP1962,Kulik75} and in the magnetic susceptibility \cite{PCScien07} is observed. 

This quantum periodicity is described enough well in large part by the relation (3) and (4) of the conventional theory \cite{Tinkham}. The persistent current (3) corresponding to the minimal energy (4) is diamagnetic at $n'\Phi_{0} < \Phi < (n'+0.5)\Phi_{0}$  and paramagnetic at $(n'+0.5) \Phi_{0} < \Phi < (n'+1)\Phi_{0}$. The maximums of the critical temperature oscillations $T _{c}(\Phi)$ are observed at $\Phi = n'\Phi_{0}$ and the minimums at $\Phi = (n'+0.5)\Phi_{0}$ \cite{LP2001} because the fractional depression of the transition temperature depends on the kinetic energy (4) $\Delta T _{c}/T _{c} \propto  -E _{k} \propto -(n\Phi_{0} - \Phi)^{2}$ \cite{Tinkham}. As a consequence the resistance $\Delta R(\Phi) \propto -\Delta T _{c}/T _{c}$ measured in the fluctuation region near the transition temperature $T \approx T _{c}$, where the resistance changes from $R = 0$ at $T < T _{c}$ to $R = R _{n}$ at $T > T _{c}$,  has the maximums at $\Phi = (n'+0.5)\Phi_{0}$ and the minimums at $\Phi = n'\Phi_{0}$ \cite{LP1962,Letter07,Kulik75}. The quantum oscillations of the magnetic susceptibility $\Delta \Phi _{Ip}$ measured in the fluctuation region near $T _{c}$ \cite{PCScien07} and the dc voltage $V_{dc}(\Phi )$ \cite{APL2016,Letter07,PLA2012,Dubonos02,Letter2003} are alternating because of their proportionality $\Delta \Phi _{Ip} = L_{f}\overline{I_{p}}$ \cite{PCScien07} and  $V_{dc}(\Phi ) \propto \overline{I_{p}}(\Phi )$ to the persistent current $\overline{I_{p}} = \int _{0}^{\Theta }dt I(t)/\Theta $ average in time ${\Theta } \gg 1/f _{sw}$, where $f _{sw}$ is the frequency of ring's switching between superconducting and normal states. The values $\Delta \Phi _{Ip}$ and $V_{dc}$ equal zero at $\Phi = n'\Phi_{0}$ and $\Phi = (n'+0.5)\Phi_{0}$  since $\overline{I_{p}} \approx (n'\Phi_{0} - \Phi)/L _{k} = 0$ at $\Phi = n'\Phi_{0}$ and $\overline{I_{p}} \approx P _{n'} (n'\Phi_{0} - \Phi)/L _{k} + P _{n'+1}[(n'+1)\Phi_{0} - \Phi]/L _{k}  = 0$ due to the equality of the current values $I _{p} (n') = -I _{p}(n'+1)$  (3), the energy $E _{k} (n') = -E _{k}(n'+1)$ (4) and thus the probability $P _{n'} = P _{n'+1}$ at $\Phi = (n'+0.5)\Phi_{0}$. 

\begin{figure}
\includegraphics{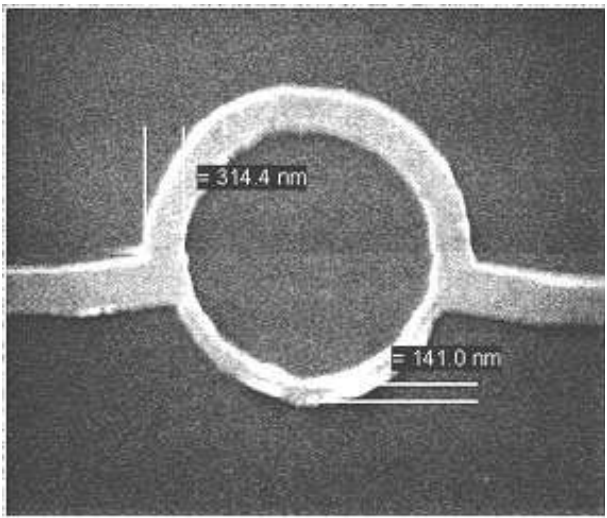}
\caption{\label{fig:epsart} The aluminum ring $r \approx 1 \mu m$ with asymmetric link-up of current leads, $l_{sh} \approx 0.7\pi r$, $l_{long} \approx 1.3\pi r$, and different width of the short $w_{sh} \approx 0.15 \ \mu m$ and long $w_{sh} \approx 0.3 \ \mu m$ segments.}
\end{figure}

Measurements of the critical current of symmetric rings corroborate also the predictions of the conventional theory \cite{Tinkham}. According to this theory the critical current of the ring with identical halves should be equal 
$$I_{c} = I_{c0} - 2|I _{p}| =  I_{c0} - 2I_{p,A}2|n- \frac{\Phi }{\Phi _{0}}| \eqno{(5)}$$
because the persistent current increases the total current in one of the ring halves, see the deducing of (5) in \cite{PCJETP07}. Here $I_{c0}$ is the critical current of the ring at the persistent current equals zero $I_{p} = 0$; $I_{p,A} =  \Phi _{0}/2L_{k}$ is the persistent current at $|n - \Phi /\Phi _{0}| = 1/2$. Experimental oscillations $I_{c}(\Phi )$ of the symmetric ring, see Fig.2 of \cite{JETP07J}, agree with (5). Measurements of the $I_{c}(\Phi )$ oscillations corroborate that the ring transform into superconducting state with the winding number $n$ corresponding to the minimum of the kinetic energy (4). The ring is switched in the normal state by the external current $|I_{ext}| > I_{c}$ at each $I_{c}$ measurement and comes back into superconducting state at $|I_{ext}| < I_{rs} < I_{c}$ \cite{PRB2014}. Each of the numerous measurements gives the same $I_{c}$ value corresponding to the same $n = n'$ at $(n' +0.5)\Phi _{0} > \Phi > (n' - 0.5)\Phi _{0}$ \cite{JETP07J,PCJETP07}. Two $I_{c}$ values corresponding to different numbers, $n = n'$ and $n = n'+1$, or  $n = n'$ and $n = n'-1$ were observed only in a specially shaped loop \cite{2statesLT34}.  

The critical current of the asymmetric ring with the different cross-sectional area of the halves $s_{w} > s_{n}$  should be anisotropic $I_{c,an} = I_{c+} - I_{c-} =  I_{p}(s_{n}/s_{w} - s_{w}/s_{n})$ \cite{PCJETP07}. Here $I_{c+}$ and $I_{c-}$ is the critical current measured at the direction of the measuring current $I_{ext}$ from left to right and from right to left; the clockwise direction of the persistent current corresponds a positive $I_{p}$ value; the top half of the ring is narrow with the cross-sectional area $s_{n}$, see Fig.1 of \cite{PCJETP07}. Measurements corroborate the anisotropy of the critical current of asymmetric rings \cite{PCJETP07}. The rectification of the ac current is observed \cite{Letter2003} due to this anisotropy \cite{PCJETP07}. But the measurements \cite{JETP07J,PCJETP07} have revealed that the reason of the anisotropy $I_{c,an}$ differs fundamentally from the expected one. The critical current of the symmetric aluminum ring with $s_{w} = s_{n}$ \cite{JETP07J} is isotropic $I_{c,an} = I_{c+} - I_{c-} = 0$ because of the equality $I_{c+}(\Phi ) = I_{c-}(\Phi ) = I_{c}(\Phi )$. Its quantum periodicity, both the theoretical one (5) and the experimental one, see Fig.2 of \cite{JETP07J}, has maximum at $\Phi = n'\Phi _{0}$ and minimum at $\Phi = (n' +0.5)\Phi _{0}$. Whereas the $I_{c+}(\Phi )$, $I_{c-}(\Phi )$ extremes of the asymmetric aluminum ring with $s_{w} \approx 2s_{n}$ are observed at $\Phi \approx (n' \pm 0.25)\Phi _{0}$ \cite{JETP07J}. These experimental quantum periodicity may be described with the relations $I_{c+}(\Phi ) \approx I_{c}(\Phi + 0.25\Phi _{0})$, $I_{c-}(\Phi ) \approx I_{c}(\Phi - 0.25\Phi_{0})$, where $I_{c}(\Phi )$ is described with (5) and observed at the measurement of the symmetric ring \cite{JETP07J}. The shift on $\approx 0.25\Phi_{0}$ in opposite direction because of the ring asymmetry provides the anisotropy $I_{c,an} = I_{c+}(\Phi ) - I_{c-}(\Phi ) \approx I_{c}(\Phi + 0.25\Phi _{0}) - I_{c}(\Phi - 0.25\Phi _{0}) \neq 0$. But the observation of the $I_{c+}(\Phi )$, $I_{c-}(\Phi )$ extremes at $\Phi \approx (n' \pm 0.25)\Phi _{0}$ is a puzzle contradicting to the theory \cite{QuStMatFoun} and the observation of the quantum periodicity in the resistance \cite{JETP07J}. The resistance $\Delta R(\Phi)$ of both symmetric and asymmetric rings has maximum at $\Phi = (n' +0.5)\Phi _{0}$ and minimum at $\Phi = n'\Phi _{0}$ \cite{JETP07J} in accord with the conventional theory \cite{Tinkham}.  

The absence of the $I_{c+}(\Phi )$, $I_{c-}(\Phi )$ jump which must be observed with the change of the winding number $n$ reveals other puzzle. Numerous observations \cite{APL2016,PCJETP07,Letter2003,Letter07,PLA2012,PCScien07} of the quantum periodicity in different parameters give evidence of the $n$ change from $n = n'$ to $n = n' + 1$ at $\Phi = (n' +0.5)\Phi _{0}$. According to the conventional theory \cite{Tinkham} the pair velocity (3a) $v$, see Fig.4, and the persistent current (3) should be inverted with this $n$ change. This inversion cannot change the critical current of the symmetric ring according to (5) and should change by jump from $I_{c} = I_{c0} - I_{p,A}(1 + s_{w}/s_{n})$ to $I_{c} = I_{c0} - I_{p,A}(1 + s_{n}/s_{w})$ the critical current of the asymmetric ring \cite{PCJETP07}. The jump $\Delta I_{c} = (s_{w}/s_{n} - s_{n}/s_{w})I_{p,A}$ should have a noticeable value $\Delta I_{c} \approx 1.5 I_{p,A}$ for the ring with $s_{w} \approx 2s_{n}$ and the amplitude of the persistent current $I_{p,A} \approx  0.2 \ mA (1 - T/T_{c})$ \cite{PCJETP07}. But measurements of asymmetric aluminum rings with $s_{w} \approx 2s_{n}$ \cite{PCJETP07,JumpAbs2006} provide reliable evidence the absence of any $I_{c}$ jump. The quantum periodicity in the critical current $I_{c+}(\Phi )$, $I_{c-}(\Phi )$ leaves no doubt that the persistent current (3) is not zero, the spectrum of the permitted states (4) is discrete, the winding number $n$ is exactly a certain number for a given $\Phi $ value and this certain number changes to 1 with the change of the magnetic flux $\Phi $ on the flux quantum $\Phi _{0}$. But the $I_{c}$ jump which could be connected with the change from $n = n'$ to $n = n' + 1$ is not observed.     

This puzzle discovered thanks to the measurements of asymmetric rings  \cite{PCJETP07,JumpAbs2006} is more conveniently to investigate using a ring with asymmetric link-up of current leads. According to the results of the measurements \cite{PCJETP07} the whole ring is switched in the normal state by jump \cite{PRB2014} when the pair velocity reaches the depairing velocity $v _{c}$ in any its segment. The external current $I_{ext} = I_{ext,l} + I_{ext,s} = qsn_{s}(v_{long} - v_{sh})$ should distribute in the ratio $I_{ext,s}/I_{ext,l} = -v_{sh}/v_{long} = l_{long}/l_{sh}$ between the short $l_{sh}$ and long $l_{long}$ segments of a homogeneous ring with equal $s$ and $n_{s}$ values along the circumference, because $v_{long}l_{long} + v_{sh}l_{sh} = 0$ at $n - \Phi /\Phi _{0} = 0$ according to the quantization condition (3a). The positive direction is taken to be from left to right for the external current $I_{ext}$, and clockwise for the persistent current $I_{p} $ (4) and the pair velocity $v_{sh}$, $v_{long}$. The short segment of the ring is situated at the bottom, see Fig. 1. Therefore $I_{ext,l} = qsn_{s}v_{long}$ and $I_{ext,s} = - qsn_{s}v_{sh}$. The external current $I_{ext} = - qsn_{s}v_{sh}(l_{sh}/l_{long} + 1) =  - qsn_{s}v_{sh}l/l_{long}$ reaches the critical value $|I_{ext}| = I_{c0} = qsn_{s}v_{c}l/l_{long}$ at $|v_{sh}| = v_{c}$ because $|v_{sh}| = |v_{long}|l_{long}/l_{sh} > |v_{long}|$ at $l_{long} > l_{sh}$, where $l = l_{sh} + l_{long} = 2\pi r$. The persistent current $I_{p}$ may both increase and decrease of the velocity $|v_{sh}|$. Consequently, it may both decrease (when the currents $I_{ext}$ and $I_{p}$ have the same direction in the short segment) and increase (when the currents $I_{ext}$ and $I_{p}$ have the opposite direction in the short segment) the critical current. Magnetic dependence of the critical current should replicate the dependence of the persistent current (3) or of the circular velocity of Cooper pairs (3a), if $|v_{sh}| > |v_{long}|$ at any value of the magnetic flux, see Fig.4. Therefore we should expect to observe the jump of the critical current because of the velocity jump at $\Phi  = (n+0.5)\Phi _{0}$ predicted by the conventional theory, see Fig.4.5 of \cite{Tinkham}.    

The values of the velocity in the short $l_{sh}$ and long $l_{long}$ segments are connected by the relation  
$$l_{long} v_{long} +  l_{sh} v_{sh} =  \frac{2\pi \hbar }{m}(n -  \frac{\Phi }{\Phi _{0}}) \eqno{(6)}$$
according to (3a). The circular persistent current $I_{p} \neq 0$ at $n - \Phi /\Phi _{0} \neq 0$ increases (or decreases) the velocity in the short segment  
$$v_{sh} =  -\frac{l_{long}}{l}\frac{I_{ext}}{qsn_{s}} + \frac{2\pi \hbar }{ml}(n -  \frac{\Phi }{\Phi _{0}}) \eqno{(7a)}$$
and decreases (or  increases) the velocity in the long segment
$$v_{long} =  \frac{l_{sh}}{l}\frac{I_{ext}}{qsn_{s}} + \frac{2\pi \hbar }{ml}(n -  \frac{\Phi }{\Phi _{0}}) \eqno{(7b)}$$
according to (6). The pair velocity reaches the critical value first in the short segment $|v_{sh}| = v_{c} > v_{long}$ at any value $|n - \Phi /\Phi _{0}| \leq 0.5$ in the ring with a strong asymmetry which is determined by the inequality $(l_{long} - l_{sh})l_{long}/l^{2} \geq  I_{p,A}/I_{c0}$ according to (7), where $I_{p,A} = qsn_{s}2\pi \hbar/2ml=  \Phi _{0}/2L_{k}$. The jump $\Delta  I_{c} = \Delta  I_{p}l/l_{long}$ of the critical current   
$$I_{c} = |\pm I_{c0} +  \frac{l}{l_{long}}I_{p}| = |\pm I_{c0} +  \frac{l}{l_{long}}I_{p,A}2(n -  \frac{\Phi }{\Phi _{0}})| \eqno{(8a)}$$
should exceed in this case the jump of the persistent current equal $\Delta  I_{p} = I_{p}(n'+1) - I_{p}(n') = ((n'+1)\Phi_{0}-\Phi )/L_{k} - (n'\Phi_{0}-\Phi )/L_{k} = \Phi_{0}/L_{k} = 2I_{p,A}$ according to (3). The critical current $I_{c0} = qsn_{s}v_{c}l/l_{long}$ of the ring at $I_{p} = 0$ is taken positive $+I_{c0}$ in (8a) when the external current $I_{ext}$ is directed from left to right and the negative value $-I_{c0}$ is taken when the current $I_{ext}$ has the opposite direction. The direction of the persistent current $I_{p}$ circulating clockwise and of the $I_{ext}$ directed from left to right is opposite in the short segment of the ring situated at the bottom, see Fig. 1. Therefore the persistent current increases the value of the critical current (8a) in this case. The direction of $I_{ext}$ and $I_{p}$ is opposite in the short segment also when the current $I_{ext}$ is directed from right to left ($-I_{c0}$ is taken in this case) and the persistent current circulates anticlockwise ($I_{p} < 0$ in this case). The $I_{c}$ jump should be smaller $\Delta I_{c} = (l_{long} - l_{sh})(I_{c0}/l_{sh} - I_{p,A}l /l_{long}l_{sh})$ in the ring with a weak asymmetry $(l_{long} - l_{sh})l_{long}/l^{2} < I_{p,A}/I_{c0}$. The critical current of such ring equal
$$I_{c} =  |\pm \frac{l_{long}}{l_{sh}}I_{c0} -  \frac{l}{l_{sh}}I_{p}| \eqno{(8b)}$$
in some region near $\Phi  = (n+0.5)\Phi _{0}$, where the pair velocity in the long segment exceeds the one in the short segment $|v_{long}| = v_{c} > |v_{sh}|$ due to the persistent current. The $I_{c}$ jump should decrease down to zero at $l_{long} = l_{sh} = l/2$ according to (8) because of the coincidence of (8a) and (8b) with (5) in this case. The ring with asymmetric link-up of current leads is handy for experimental investigation of the magnetic dependence of the persistent current due to the linear dependence of its critical current $I_{c}(\Phi )$ in the region near $\Phi  = n\Phi _{0}$ predicted by (8a). (The linear dependence turns into the maximum at $l_{long} = l_{sh} = l/2$ according to (5)). Measurements of the $I_{c}(\Phi )$ dependence in this region give both $I_{c0}$ and $I_{p,A}$ values.

\begin{figure}
\includegraphics{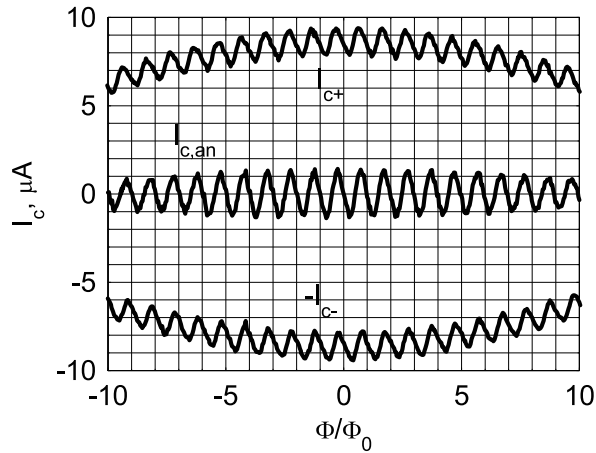}
\caption{\label{fig:epsart} The quantum periodicity in the anisotropy of the critical current $I_{c,an} = I_{c+}(\Phi ) - I_{c-}(\Phi )$ and in the critical current $I_{c+}(\Phi )$, $I_{c-}(\Phi )$ measured in the opposite directions on the aluminum ring $r \approx 1 \mu m$ with asymmetric link-up of current leads, $l_{sh} \approx 0.7\pi r$, $l_{long} \approx 1.3\pi r$, and different width of the short $w_{sh} \approx 0.15 \ \mu m$ and long $w_{long} \approx 0.3 \ \mu m$ segments at the temperature $T \approx  1.420 \ K \approx  0.93T_{c}$.}
\end{figure}

\section{Experimental Details}
 \label{}

We used two aluminum rings and one tantalum ring with asymmetric link up of current leads for the experimental investigations of the $I_{c}$ jump. The radius, the length of the short and long segments of the aluminum rings were $r \approx 1 \ \mu m$, $l_{sh} \approx  0.7\pi r$, $l_{long} \approx  1.3\pi r$, and of the tantalum ring $r \approx 0.5 \ \mu m$, $l_{sh} \approx 0.6\pi r$, $l_{long} \approx 1.4\pi r$. The width of the $l_{sh}$ and $l_{long}$ segments of one of the aluminum rings and the tantalum ring was the same $w_{sh} \approx w_{long} \approx 0.15 \ \mu m$, whereas of the other aluminum ring $w_{sh} \approx 0.15 \ \mu m$ and $w_{sh} \approx 0.3 \ \mu m$, Fig.1. The aluminum nano-structures were fabricated by the lift-off method by depositing a thin aluminum film $d \approx 10 \ nm$ thick onto a $Si/SiO_{2}$ substrate. The technology of the fabrication of the tantalum nano-structures from epitaxial Ta film with $d \approx 60 \ nm$ was described earlier \cite{TaTech}. The resistance of the ring in the normal state, the resistance ratio and the superconducting transition temperature were $R_{n} \approx  60 \ \Omega $,  $R(300K)/R(4.2K) \approx 1.6$, $T_{c} \approx  1.46 \ K$ of the aluminum ring with $w_{sh} \approx w_{long}$; $R_{n} \approx  41 \ \Omega $,  $R(300K)/R(4.2K) \approx 1.5$, $T_{c} \approx  1.52 \ K$ of the aluminum ring with $w_{sh} < w_{long}$; $R_{n} \approx  115 \ \Omega $,  $R(300K)/R(4.2K) \approx 1.7$, $T_{c} \approx  2.83 \ K$ of the tantalum ring. 

\begin{figure}
\includegraphics{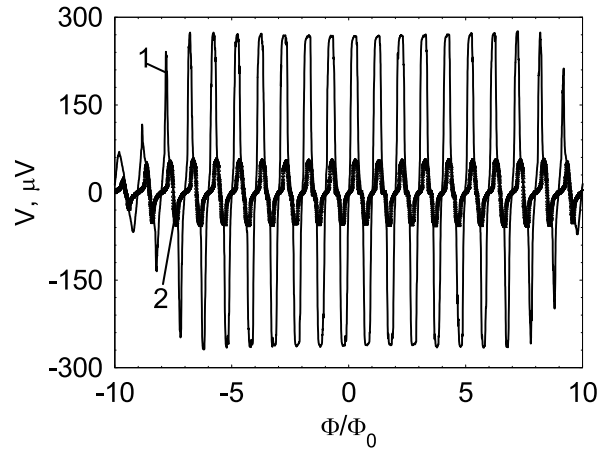}
\caption{\label{fig:epsart} The quantum periodicity in the dc voltage with the maximum amplitude  $V_{A,max}\approx 270 \ \mu V$ and $V_{A,max}\approx 0.55 \ \mu V$ induced on the aluminum ring $r \approx 1 \mu m$ with $l_{sh} \approx 0.7\pi r$, $l_{long} \approx 1.3\pi r$ and $w_{sh} \approx w_{long} \approx 0.15 \ \mu m$ by the external ac current $I_{ext} = I _{A} \sin (2\pi f)$ with the frequency $f = 30 \ Hz$ and the amplitude $I_{A,max} \approx 21 \ \mu A$ and  $I_{A,max} \approx 4.8 \ \mu A$ at the temperature $T \approx  1.288 \ K \approx  0.882T_{c}$ and $T \approx  1.394 \ K \approx 0.955T_{c}$. The voltage scale of the dependence (2) was magnified 100 times.}
\end{figure}

The measurements were carried out by a fourterminal method. We used 4He as the cooling agent and pumping of helium allowed lowering the temperature down to 1.19 K. The temperature was determined with the help of a calibrated thermistor ($R(300K)=1.5 \ k\Omega $) with an excitation current of $0.1 \ \mu A$. The dependences $I_{c+}(B)$ and $I_{c-}(B)$ of the critical current on the magnetic field were found from the periodic (10 Hz) current-voltage characteristics in a slowly ($\sim 0.01$ Hz) varying magnetic field $B_{sol}$ according to the following algorithm: (i) the superconducting state of the structure was verified; (ii) after the threshold voltage (set above the pickup and noise level of the measuring circuit and determining the lowest measurable critical current) was exceeded, the magnetic field and critical current were measured with a delay of about $30 \ \mu s$. Thus, the critical current in the positive ($I_{c+}$) and negative ($I_{c-}$) directions of the external current $I$ were measured in sequence. The magnetic field $B$ perpendicular to the sample plane was produced by a copper coil. The measured quantities were recorded as functions of the current $I_{sol}$ in the coil. The magnitude of the magnetic field induced by the current in the coil was determined from the calibration $B_{sol} = k_{sol}I_{sol}$ with $k_{sol} \approx  129 \ G/A$ found with the use of a Hall probe. The measurement of the critical currents in opposite directions allowed us to determine the external magnetic field $B = B_{sol}+B_{res}$. Since simultaneous change of the direction of the total external magnetic field $B$ and the external current is equivalent to the rotation by $180^{0}$, one has $I_{c+}(B) = I_{c+}(B_{sol}+B_{res}) = I_{c-}(-B) = I_{c-}(-B_{sol}-B_{res})$. The residual magnetic field $B_{res} \approx  0.1 \ G$ thus determined corresponds to the flux $SB_{res} = \pi r^{2}B_{res} \approx  0.02\Phi _{0}$ in the ring with the radius $r \approx   1 \ \mu m$. The sine external current $I_{ext} = I _{A} \sin (2\pi f )$ was used for the observation of the rectification effect. According to the previous measurements \cite{Letter2003} the amplitude $V_{A}$ of the rectified voltage $V_{dc}(\Phi )$ depends on the amplitude $I _{A}$ and does not depend on the frequency $f$ of the ac current at least up to $1 \ MHz $.

\begin{figure}
\includegraphics{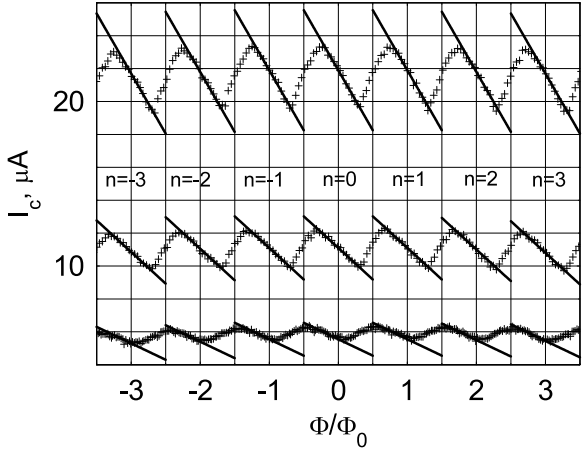}
\caption{\label{fig:epsart} Comparison of the experimental quantum periodicity in the critical current (crosses) measured on the aluminum ring with $l_{sh} \approx 0.7\pi r$, $l_{long} \approx 1.3\pi r$ and $w_{sh} \approx w_{long} \approx 0.15 \ \mu m$ measured at the temperature $T \approx 1.284 \ K \approx  0.88T_{c}$, $T \approx 1.345 \ K \approx  0.92T_{c}$ and $T \approx  1.393 \ K \approx  0.95T_{c}$ with the theoretical prediction (8a) (lines) corresponding to the following values of the critical current and the amplitude the persistent current at $\Phi = 0$: $ I_{c0} \approx  22 \ \mu A$, $I _{p,A} \approx  3 \ \mu A$ at $T \approx  0.88T_{c}$; $I_{c0} \approx  11 \ \mu A$, $I _{p,A} \approx  2 \ \mu A$ at $T \approx  0.92T_{c}$; $I_{c0} \approx  5.5 \ \mu A$, $I _{p,A} \approx  1 \ \mu A$ at $T \approx  0.95T_{c}$. The top drawing shows the magnetic dependence of the circular velocity of Cooper pairs $v = \Phi /\Phi _{0} - n$, see (3a), at $I_{ext} = 0$, at the winding number $n$ corresponds to the minimal kinetic energy (4), see Fig.4.5 of \cite{Tinkham}.}
\end{figure}

\section{Experimental Results and Discussion }
 \label{}
 
The critical current of the ring with asymmetric link up of current leads $l_{sh} < l_{long}$, as well as of the asymmetric ring with $s_{w} > s_{n}$  \cite{PCJETP07}, should be anisotropic $I_{c,an} = I_{c+}(\Phi ) - I_{c-}(\Phi ) \neq 0$ at $I_{p} \neq 0$ because of the change of the $I_{c0}$ sign in (8) with the $I_{ext}$ direction change: $I_{c,an} = I_{c+}(\Phi ) - I_{c-}(\Phi ) = |+I_{c0} +  (l/l_{long})I_{p}| - |-I_{c0} +  (l/l_{long})I_{p}| = (2l/l_{long})I_{p}$ for the ring with the strong asymmetry and $I_{c0} >  (l/l_{long})|I_{p}|$. Measurements of the critical current have corroborated the $I_{c}$ anisotropy, Fig.2, of all our rings. The rectification effect observed, Fig.3, due to this anisotropy does not differ from the one observed at measurements of asymmetric rings \cite{PCJETP07,Letter2003}. The rectified voltage $V_{dc}(\Phi )$ appear when the current amplitude $I _{A}$ exceeds the lowest value of the critical current $I_{c}(\Phi ,T)$ at the temperature $T$ of measurement. The $V_{dc}(\Phi )$ amplitude $V _{A}$ increases with the $I _{A}$ increase up to a maximum value $V_{A,max}$ at $I _{A} = I_{A,max}$ and decreases thereafter, see Fig.6 of \cite{Letter2003}  and Fig.12 of \cite{PCJETP07}. The value $Eff_{Re} = \surd 2V_{A,max}/I_{A,max}R_{n}$ defines the rectification efficiency \cite{PLA2012}. This value is high $Eff_{Re} \approx 0.3$ at low temperature and decreases near $T_{c}$ according to the measurements of both the asymmetric rings, see Fig.15 of \cite{PCJETP07}, Fig.6 of \cite{Letter07} and the rings with asymmetric link up of current leads, Fig.3. According to the experimental results presented on Fig.3, $Eff_{Re} \approx  0.3$ ($V_{A,max} \approx  270 \ \mu V$ at $I_{A,max} \approx  21 \ \mu A$) at $T \approx  0.882T_{c}$ and  $Eff_{Re} \approx  0.003$ ($V_{A,max} \approx  0.55 \ \mu V$ at $I_{A,max} \approx  4.8 \ \mu A$) at $T \approx  0.955T_{c}$.

\begin{figure}
\includegraphics{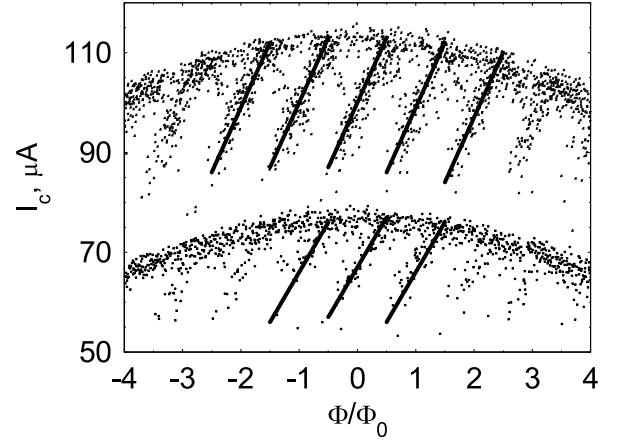}
\caption{\label{fig:epsart} Comparison of the experimental quantum periodicity in the critical current (points) of the tantalum ring with $r \approx 0.5 \ \mu m$, $l_{sh} \approx 0.6\pi r$, $l_{long} \approx 1.4\pi r$ and $w_{sh} \approx w_{long} \approx 0.15 \ \mu m$ measured at the temperature $T \approx  2.38 \ K \approx 0.84T_{c}$ and $T \approx  2.48 \ K \approx 0.88T_{c}$ with the theoretical prediction (8a) (lines) corresponding to the following values of the critical current and the amplitude the persistent current at $\Phi = 0$: $ I_{c0} \approx  100 \ \mu A$, $I _{p,A} \approx  9 \ \mu A$ at $T \approx  0.84T_{c}$; $I_{c0} \approx  67 \ \mu A$, $I _{p,A} \approx  7 \ \mu A$ at $T \approx  0.88T_{c}$.}
\end{figure} 

The critical current of the aluminum rings varies linearly with magnetic field in accordance with the theoretical prediction (8a) in the regions close to the integer numbers of the magnetic flux $\Phi \approx n'\Phi _{0}$, Fig.4. This agreement between experiment and theory allows to measure the critical current $I_{c0} = I_{c}(\Phi = n'\Phi _{0})$ and the amplitude $I_{p,A} =  \Phi _{0}/2L_{k}$ of the persistent current  $I_{p} = I_{p,A}2(n -\Phi /\Phi _{0})$ using (8a). The $|I_{c0}|$ value decreases with the $|n'|$ increase, Fig.2,5, because of superconductivity depression in the strips with the non-zero width $w_{sh}$, $w_{long}$ by the magnetic field $B$. Measurements of the oscillations $I_{c+}(\Phi )$,  $I_{c-}(\Phi )$ at different temperatures allow to define the temperature dependence of $I_{c0}(T)$ and $I_{p,A}(T)$ at $\Phi = 0$. These dependence may be described with the relations $I_{c0} =  I _{c}(T=0)(1 - T/T _{c})^{3/2}$ and $I _{p,A} = I _{p,A}(T=0)(1 - T/T _{c})$ in accordance with the theoretical prediction \cite{Tinkham} with $T _{c} = 1.46 \ K$, $I _{c}(T=0) = 500 \ \mu A$, $I _{p,A}(T=0) = 22.5 \ \mu A$ for the aluminum ring with $w_{sh} = w_{long}$ and $T _{c} = 1.52 \ K$, $I _{c}(T=0) = 720 \ \mu A$ , $I _{p,A}(T=0) = 36 \ \mu A$ for the one with $w_{sh} < w_{long}$. According to the conventional theory \cite{Tinkham} $I _{p,A}/I _{c0} = (1+l_{sh}/l_{long})^{-1}(\surd 3\xi (T)/2r) $ depends on the ratio of the correlation length $\xi (T)$ and the ring radius $r$ since $I _{p,A} = sqn_{s}\hbar /2mr$ and $I _{c0} = sqn_{s}(1+l_{sh}/l_{long})v_{c}= sqn_{s}(1+l_{sh}/l_{long})(\hbar /m\surd 3\xi (T)$ for the ring with $s_{long} = s_{sh} = s$. The experimental value $I _{p,A}/I _{c0}  \approx 0.05(1 - T/T_{c})^{-1/2}$ for the ring with $r \approx 1 \ \mu m$ corresponds to the theoretical prediction at the expected value of the aluminum correlation length $\xi (T) = 0.1 \ \mu m \ (1 - T/T_{c})^{-1/2} $.  

The quantum periodicity in the critical current and its anisotropy, Fig.2, leaves no doubt that the persistent current (3) is not zero and the spectrum of the permitted states (4) is strongly discrete. These observations could not be possible if the density of Cooper pairs $n _{s}$ vanishes at any point of the ring, i.e. $1/\overline{(n _{s})^{-1}} = 0$,  in the measurement process of the critical current. The density $n _{s}$ can vary only in the limits from $n_{s} = n_{s0}$ to $n_{s} = 2n_{s0}/3$ at $|I_{ext}| \leq I_{c}$, according to the conventional theory \cite{Tinkham}. Therefore the relations (8) obtained for a uniform density, i.e. for $1/\overline{(n _{s})^{-1}} = n _{s}$, is valid for the description of the quantum periodicity in the critical current $I_{c}(\Phi)$. The observation of the linear dependence $I_{c} = I_{c0} +  (l/l_{long})I_{p,A}2(n - \Phi /\Phi _{0})$ near $\Phi \approx n'\Phi _{0}$, Fig.4, testifies against a dependence of the density $1/\overline{(n _{s})^{-1}} > 0$ on the magnetic flux $\Phi $. The dependence $I_{c}(\Phi)$ replicates in this region the velocity dependence of Cooper pairs in the short segment $v_{sh}(\Phi)$ (7a) at a constant value $I_{ext} $, see Fig.4. The observation of the linear dependence $I_{c}(\Phi)$ near $\Phi \approx n'\Phi _{0}$ with different $n' = -3, -2, -1, 0, 1, 2, 3$, Fig.4, leaves no doubt that the winding number $n$ must change with $\Phi $: $n = n'$ near $\Phi \approx n'\Phi _{0}$. The inversion of the circular component (3a) of the velocity $v_{sh}$ (the second term of (7a)) with the change of the winding number from $n = n'$ to $n = n'+1$ must result to the jump of the velocity in the short segment $v_{sh}$ (7a), of the persistent current (3) and of the critical current (8) due to a non-zero density $1/\overline{(n _{s})^{-1}} > 0$. 

Our rings satisfy the criterion of strong asymmetry $(l_{long} - l_{sh})l_{long}/l^{2} \approx 0.2 \geq  I_{p,A}/I_{c0} \approx 0.05(1 - T/T_{c})^{-1/2}$ at $T < 0.93T_{c}$. The $I_{c}$ jump should be equal  $\Delta  I_{c} = 2I_{p,A}l/l_{long}$ in this temperature region. But no jump is observed at measurement of the critical current of aluminium rings with asymmetric link-up of current leads, Fig.2, Fig.4, as well as of aluminium asymmetric rings \cite{PCJETP07,JumpAbs2006}. The experimental data deviate from the theoretical dependence close to $\Phi \approx (n'+0.5)\Phi _{0}$, so that a smooth change is observed instead of the jump. This smooth change is a puzzle because the winding number $n$ in (8) must be integer number. According to (8a) the critical current at $\Phi = n'\Phi _{0}$ should be equal the one without the persistent current $I_{c} = I_{c0}$ because $I_{p} = 0$ at $n = n'$ in this case, according to (3). But the state with $I_{p} = 0$ and $1/\overline{(n _{s})^{-1}} > 0$ is forbidden at $\Phi = (n'+0.5)\Phi _{0}$. Therefore the measurement of the critical current $I_{c} = I_{c0}$ at  $\Phi = (n'+0.5)\Phi _{0}$, Fig.4, may mean an observation of the forbidden state.
   
Could the lack of the $I_{c}$ jump to be a puzzle observed at measurements only aluminum rings? We have made first measurements of a tantalum ring with asymmetric link-up of current leads, Fig.5. The dispersion of the measured $I_{c}$ values, Fig.5, exceeds appreciably the one observed at measurements of aluminium ring, Fig.4. Nevertheless groups of points demonstrate the quantum periodicity, Fig.5, which can be described with the theoretical prediction (8a). This periodicity as the function of magnetic field with the period $B_{0} \approx 0.0025 \ T$ corresponds to the flux quantum $B_{0}S = B_{0}\pi r^{2} \approx \Phi _{0}$ inside the tantalum ring with $r \approx 0.5 \ \mu m$. Groups of experimental points are concentrated near the $I_{c}$ values predicted by (8a) when $I_{c0} \approx  100 \ \mu A$, $I _{p,A} \approx  9 \ \mu A$ at $T \approx  0.84T_{c}$ and $I_{c0} \approx  67 \ \mu A$, $I _{p,A} \approx  7 \ \mu A$ at $T \approx  0.88T_{c}$. The value of the ratio $I _{p,A}/I _{c0} = (1+l_{sh}/l_{long})^{-1}(\surd 3\xi (T)/2r) \approx 0.035(1 - T/T_{c})^{-1/2}$ corresponds to the tantalum correlation length $\xi (T) = 0.03 \ \mu m \ (1 - T/T_{c})^{-1/2} $ at the ring radius $r \approx 0.5 \ \mu m$. The first measurements of the tantalum ring give a hope for the possibility of the $I_{c}$ jump. But numerous experimental points near the maximum value $I_{c} = |I_{c0}| + I_{p,A}l/l_{long}$ predicted with (8a) force to doubt in the agreement between theory and experiment. 

\section{Conclusion}
We present the first measurements of superconducting rings with asymmetric link-up of current leads in order to draw reader's attention, first of all experimenters, on the opportunity of the experimental investigations of the problem of the winding number change with the help of such rings. These investigations may have both fundamental and practical importance. The quantization (1) and the jump between the permitted states with different values of the winding number is basis of the canonical description of many quantum phenomena, for example, of the quantum periodicity in various variables. The jump of the persistent current with the $n$ change at $\Phi \approx (n'+0.5)\Phi _{0}$ is observed at measurement of the flux qubit (quantum bit), i.e. superconducting loop with three Josephson junctions \cite{1Shot02}. The jump of the critical current with the $n$ change was observed at measurement of two phase-coupled superconducting aluminum rings \cite{JETP2011}. But the jump is not  observed at measurements of the critical current of aluminum rings, asymmetric $s_{n} < s_{w}$ \cite{PCJETP07,JumpAbs2006} and with asymmetric link-up of current leads  $l_{sh} < l_{long}$, Fig.4. It is puzzle which should be solved with the help of experimental and theoretical investigations. The jump of the critical current may be used for the designing of a simple superconducting quantum interference device without Josephson junctions \cite{Letter2014}. We cannot explain why the results of the measurements of the critical current of Ta and Al rings look so different. It may be, our first measurements of Ta ring are not enough accurate. The quality of these measurements may be improve. Future measurements should answer the question: "Are there intrinsic differences between Ta and Al rings or the absence of the jump of the critical current is the puzzle intrinsic for all superconducting rings with asymmetric link-up of current leads?" 

This work has been supported partly by the Russian Science Foundation, Grant No. 16-12-00070.


\end{document}